\documentclass[12pt]{iopart}

\usepackage{graphicx}
\usepackage{epsfig}  
\begin{document}

\title[Transition temperature and EoS from lattice QCD]{Transition temperature and the equation of state from lattice QCD, Wuppertal-Budapest results}

\author{Szabolcs Borsanyi$^1$, Gergely Endrodi$^2$, Zoltan Fodor$^{1,3}$, Christian Hoelbling$^1$, Sandor Katz$^3$, Stefan Krieg$^{1,4}$, Claudia Ratti$^{5,6}$ and Kalman Szabo$^1$}

\address{$^1$  
Dep. of Physics, Wuppertal University, Gaussstr. 20, D-42119 Wuppertal, Germany}
\address{$^2$  
Inst. f\"ur Theoretische Physik, Regensburg University, D-93040 Regensburg, Germany.}
\address{$^3$ Inst. for Theoretical Physics, E\"otv\"os University, P\'azm\'any 1, H-1117 Budapest, Hungary} 
\address{$^4$ J\"ulich Supercomputing Centre, Forschungszentrum J\"ulich, D-52425
J\"ulich, Germany}
\address{$^5$ Dip. di Fisica Teorica, Universit\`a di Torino, via Giuria 1, I-10125 Torino, Italy}
\address{$^6$ INFN, Sezione di Torino}
\begin{abstract}
The QCD transition is studied on lattices up to $N_t=16$. The chiral condensate is presented as a function of the temperature, and the corresponding transition temperature is extracted. The equation of state is determined on lattices with $N_t=6,8,10$ and at some temperature values with $N_t=12$. The pressure and the trace anomaly are presented as functions of the temperature in the range 100 ...1000 MeV . Using the same configurations we determine the continuum extrapolated 
phase diagram of QCD on the $\mu-T$ plane for small to moderate 
chemical potentials. Two transition lines are defined with two 
quantities, the chiral condensate and the strange quark number 
susceptibility.

\end{abstract}

\maketitle

\section{Introduction}
The study of QCD thermodynamics is receiving increasing interest in recent years. 
A systematic approach to determine the properties 
of the deconfinement phase transition is through lattice QCD.  Lattice simulations indicate that the transition at vanishing 
chemical potential is merely an analytic crossover
\cite{8}.
Some interesting quantities that can be extracted from lattice simulations are the transition temperature $T_c$, the QCD equation of state and, for small chemical potentials, the phase diagram in the $\mu-T$ plane: we review the results on these observables that have been obtained by our collaboration using the staggered stout action 
with physical light and strange quark masses, thus $m_s /m_{ud}\simeq28$ \cite{6,7}.
For all details we refer the reader to Refs. \cite{Borsanyi:2010bp,Borsanyi:2010cj,Endrodi:2011gv}.

\begin{figure}
\begin{minipage}{.48\textwidth}
\parbox{6cm}{
\scalebox{.52}{
\includegraphics{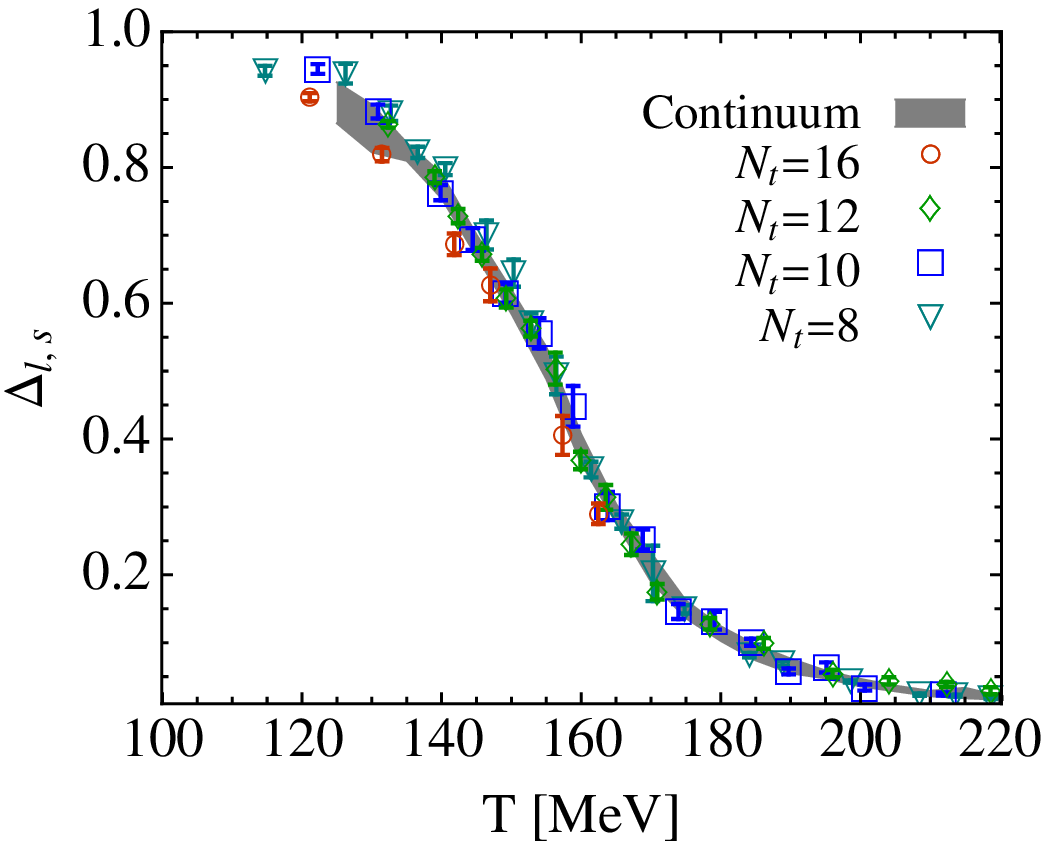}\\}}
\end{minipage}
\hspace{-.4cm}
\begin{minipage}{.48\textwidth}
\parbox{6cm}{
\scalebox{.54}{
\epsfig{file=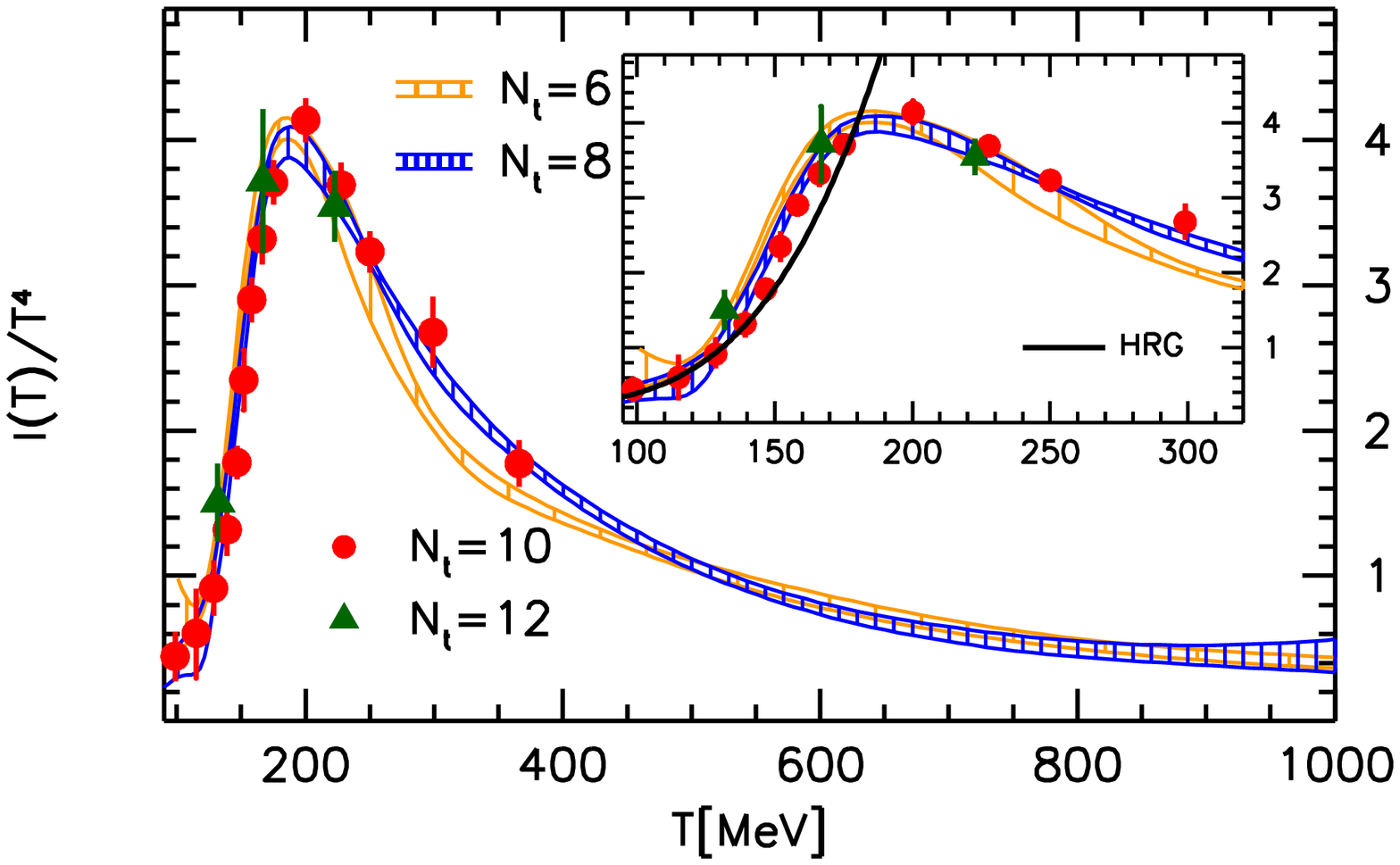,height=9cm,bb=18 360 592 718}\\}}
\end{minipage}
\caption{
Left: Subtracted chiral condensate $\Delta_{l,s}$ as a function of the 
temperature. The gray band is the continuum result of our collaboration, obtained with the 
stout action. Right: The trace anomaly normalized by $T^4$ as a function of $T$ on $N_t = 6,~ 8,~ 10$ and 12 lattices. The inset shows a comparison with the results of the Hadron Resonance Gas model, including resonances from the Particle Data Book up to 2.5 GeV mass.}
\label{fig3}
\end{figure}
\section{QCD transition temperature and Equation of State}
We present here the results for the chiral condensate, and extract the value of $T_c$ associated to this observable;
for the values of $T_c$ obtained from other observables, which reflects the nature of the
crossover transition, we refer the reader to Ref. \cite{Borsanyi:2010bp}.
The chiral condensate is defined as
$\langle\bar{\psi}\psi\rangle_q=T\partial\ln Z/(\partial m_q V)$ for q=u,d,s.
It is an indicator for the remnant of the chiral transition, since it
rapidly changes around $T_c$. We calculate the quantity
$\Delta_{l,s}$, which is defined as
$[{\langle\bar{\psi}\psi\rangle_{l,T}-{m_l}/{m_s}\langle\bar{\psi}\psi\rangle_{s,T}}]/
 [{\langle\bar{\psi}\psi\rangle_{l,0}-{m_l}/{m_s}\langle\bar{\psi}\psi\rangle_{s,0}}]$ for l=u,d.
Since the results at different lattice spacings are essentially 
on top of each other, we connect
them to lead the eye (see the left panel of Fig. \ref{fig3}).
The value of $T_c$ that we obtain from the inflection point of this observable is $T_c=157(3)(3)$.

Next we present our results regarding the equation of state; in the
right panel of Figure \ref{fig3}, the $T$ dependence of the interaction measure
is shown for the $2+1$ flavor system.
We have results at
four different lattice spacings. Results
show essentially no dependence on ``$a$", they all lie on top of
each other.  Only the coarsest $N_t=6$ lattice shows some deviation
around $\sim 300$ MeV. On the same figure, we zoom in to the
transition region. Here we also show the results from the Hadron Resonance Gas
model: a good agreement with the lattice results is found up to $T\sim 140$ MeV.

In order to obtain the pressure, we determine its partial 
derivatives with respect to the bare lattice parameters. 
 $p$ is then rewritten as a
multidimensional integral along a path in the space of bare 
parameters. To obtain the EoS for various $m_\pi$,
we simulate for a wide range of bare parameters on the plane of $m_{u,d}$ and $\beta$
($m_s$ is fixed to its physical value).  Having
obtained this large set of data we generalize the integral method
and include all possible integration paths into the analysis 
\cite{Borsanyi:2010cj,Endrodi:2010ai}.
 We remove the additive divergence of $p$ by subtracting the same observables
measured on a lattice, with the same bare parameters but at a different $T$
value. Here we use lattices with
a large enough temporal extent, so it can be regarded as $T=0$.
\vspace{-.2cm}
\section{The QCD phase diagram at nonzero quark density}
\vspace{-.2cm}
We provided results for the transition temperatures $T_c$ at vanishing 
chemical potential ($\mu=0$). Now we move out to the $\mu\neq$0 plane 
(for the details see \cite{Endrodi:2011gv}). As one increases $\mu$ the 
transition temperature $T_c(\mu^2)$ decreases. Let us parameterize the 
transition line in the vicinity of the vertical $\mu$=0 axis as
$T_c(\mu^2)=T_c(1-\kappa\cdot \mu^2/T_c^2).$

In order to determine the transition temperature as a function of $\mu$ 
we use two quantities which are monotonic in the transition region and 
do not depend on $\mu$ for zero or infinite temperatures. The transition 
temperature is defined as the temperature value at which these 
observables take their value as given by the inflection points of the 
curves at $\mu$=0.

The two observables we use are the renormalized chiral condensate and 
the normalized strange quark number susceptibility. In order to measure 
the $\mu$ dependence of these quantities we apply reweighting. Since our
lattices are quite large the full reweighting method is quite expensive.
Therefore, we truncate the $\mu$ dependence of the weights at $\mu^2$ 
order (this truncation of the original \cite{Fodor:2001au,Fodor:2004nz} 
method is usually called the Taylor method; for a recent application see
\cite{Kaczmarek:2011zz}).

The strange quark number susceptibility is the second derivative of the 
partition function with respect to the strange chemical potential. It 
needs no renormalization, since it is related to a conserved current. It 
is useful to normalize it by $T^2$, which provides a dimensionless 
combination. It is easy to see that at $T=0$ its value is 0, whereas for 
infinitely large temperatures it approaches the 
Stefan-Boltzmann limit of 1.

For the chiral condensate we apply here a slightly different
renormalization prescription than what was used for the determination
of $T_c$. We cancel the additive divergences by subtracting the $T=0$
contribution, while the multiplicative
divergence due to the derivative with respect to the mass can be 
eliminated with a multiplication by the bare quark mass. Then, in order 
to have a dimensionless combination the whole expression can be divided 
by the fourth power of some dimensionful mass scale. In this work we use 
the T=0 pion mass for the normalization. This observable has also $\mu$ 
independent limiting values at zero and at infinitely high temperatures 
(at T=0 this is true for chemical potetials smaller than the baryon mass, 
well within our applicability region).

\begin{figure}
\begin{minipage}{.48\textwidth}
\parbox{6cm}{
\scalebox{.54}{
\epsfig{file=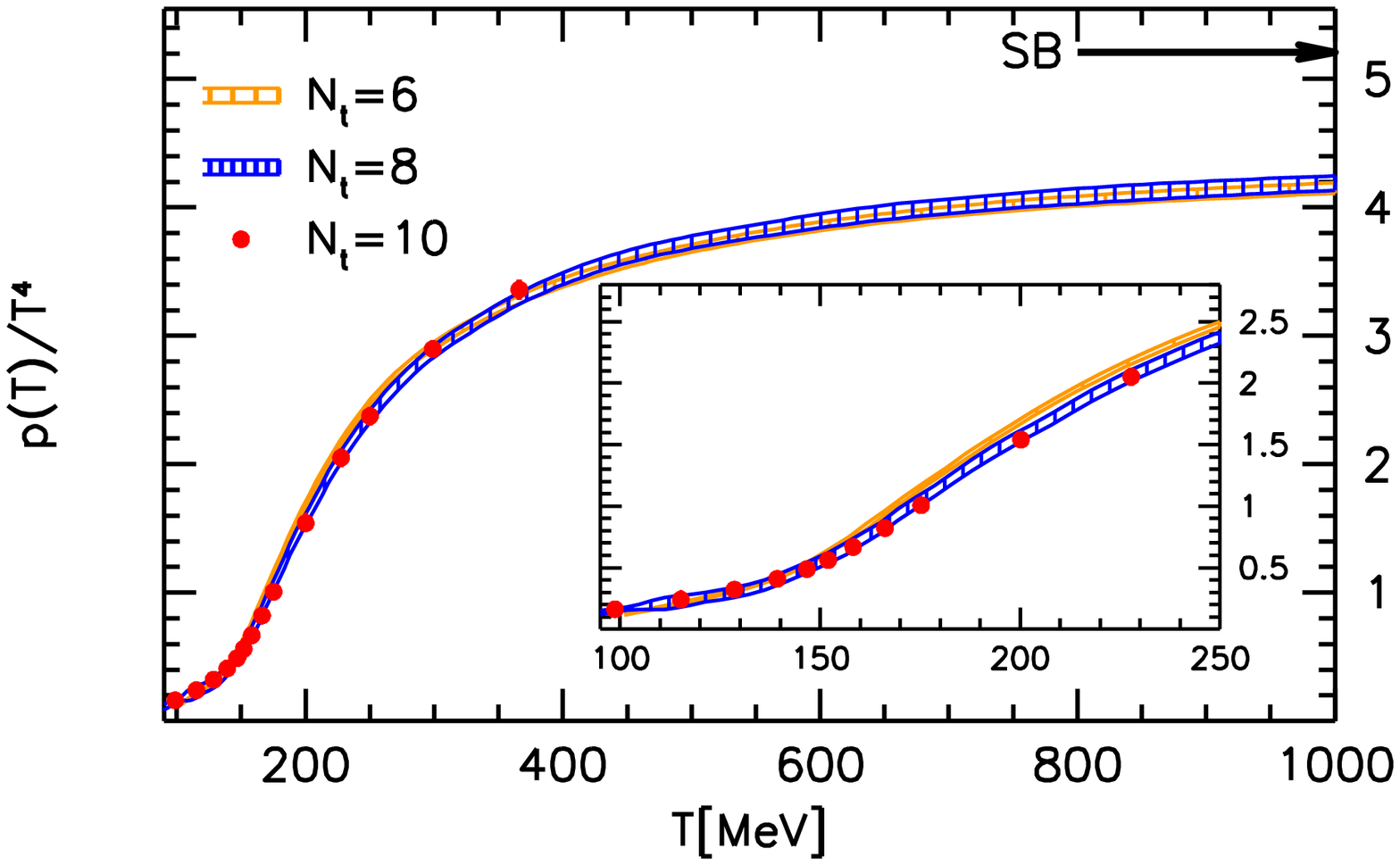,height=9cm,bb=18 360 592 718}\\}}
\end{minipage}
\hspace{.6cm}
\begin{minipage}{.42\textwidth}
\parbox{6cm}{
\scalebox{.32}{
\includegraphics{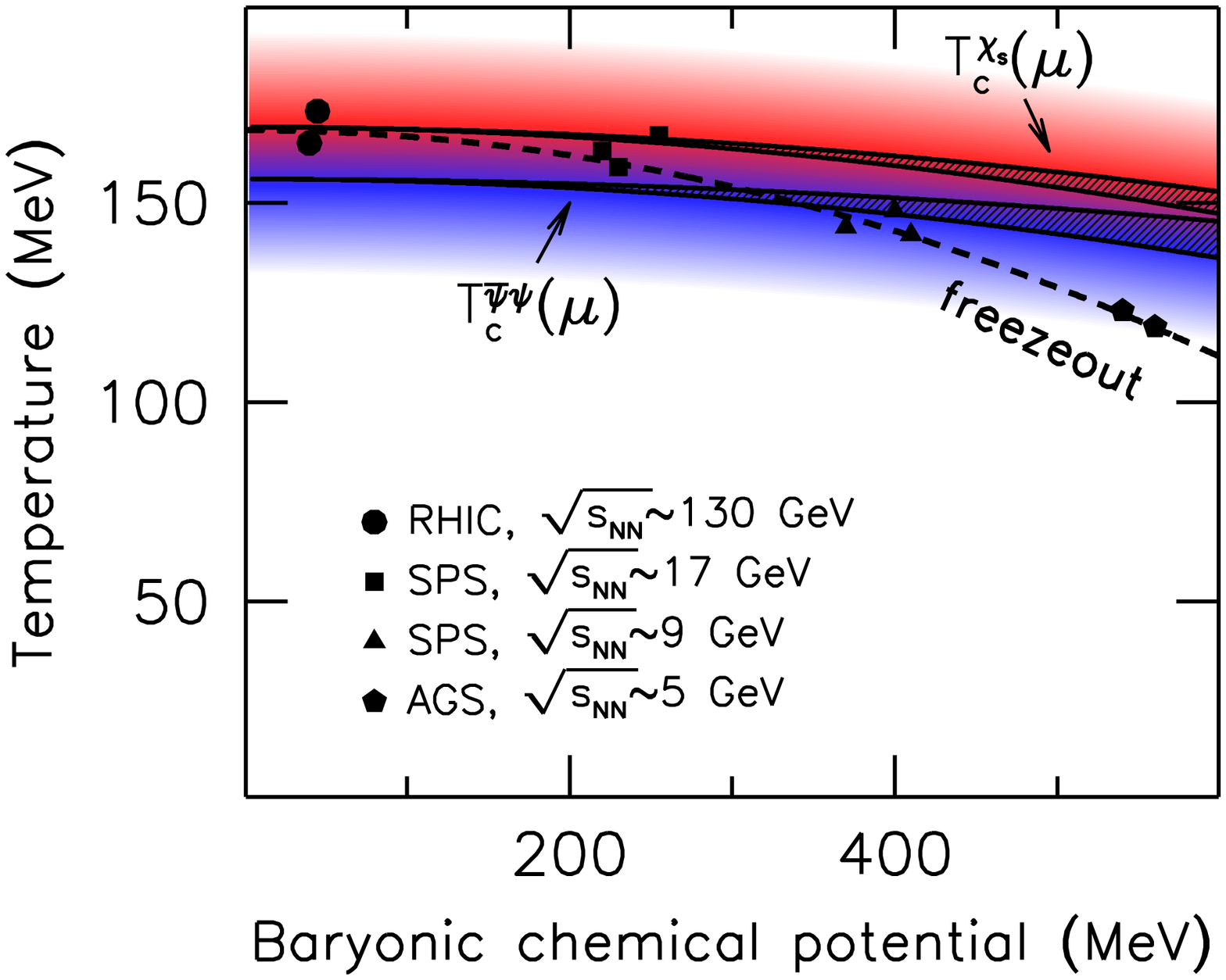}}}\\
\end{minipage}
\caption{
Left: the pressure normalized by $T^4$ as a function of $T$ on $N_t = 6,~ 8$ and 10 lattices. Right: The crossover transition between the ÔcoldÕ and ÔhotÕ phases is 
represented by the coloured area (blue and red correspond to the 
transition regions obtained from the chiral condensate and the strange 
susceptiblity, respectively). The lower solid band shows the result for 
$T_c(\mu)$ defined through the chiral condensate and the upper one 
through the strange susceptibility. The width of the bands represent the 
statistical uncertainty of $T_c(\mu)$ for the given $\mu$ coming from 
the error of the curvature for both observables. The dashed line is the 
freeze-out curve from heavy ion experiments. The center of mass energies
of these experiments are also shown.}
\label{curv}
\end{figure}

Our final result is shown in the right panel of Figure \ref{curv}. The crossover regionÕs extent 
changes little as the chemical potential increases, and within it two 
definitions give different curves for $T_c(\mu)$. It is useful to compare the 
whole picture to the freeze-out curve which summarizes 
experimental results on the T--$\mu$ points where hadronization of the 
quark-gluon plasma was observed. This curve is expected to lie in the 
interior of the crossover region, as is indicated by our results as 
well.
\vspace{-.4cm}
\section*{Acknowledgements}
\vspace{-.2cm}
Work supported in part by the EU grant  (FP7/2007-2013)/ERC no. 208740. The work of C. R. is supported by funds provided by the Italian 
Ministry of Education, Universities and Research under the Firb Research Grant 
RBFR0814TT. 

\end{document}